# Nanoscale Inclined Four-electrode Ion Traps: Finite-Element Modelling, Floquet Stability, and Surface-Potential Response

**Nahiyan Archa[1], Abhinand P.[1], Ahammed Shabeeb[2], Nikhil Kumar[3],***

[1] Department of Chemistry, National Institute of Technology Calicut, India
[2] Department of Physics, National Institute of Technology Calicut, India
[3] Department of Electronics and Communication Engineering, National Institute of Technology Calicut, India
**Corresponding author: nikhilkumarcs@nitc.ac.in*

## Abstract

The development of miniaturized trapped-ion systems requires electrode architectures capable of providing controlled radio-frequency (RF) confinement while accounting for the influence of nanoscale electrode features and surface-induced perturbations. Here, we present a finite-element investigation of a proposed inclined four-blade RF ion-trap architecture for $^{171}Yb^{+}$, integrating pseudopotential characterization, matrix Mathieu analysis, Floquet stability assessment, and localized surface-potential response modelling. The architecture incorporates nanoscale exposed electrode-face features with characteristic dimensions of approximately 99-150 nm, within a configuration having an opposing-electrode separation of 20 µm. At an electrode inclination of 6°, an RF amplitude of 150 V, and a drive frequency of 15 MHz, the calculated pseudopotential exhibits directional in-plane barriers of 3.430 and 3.389 meV along the two principal diagonal paths. The full RF-potential Hessian, including off-diagonal coupling terms, is employed to construct the matrix Mathieu equation. Numerical propagation of the associated state-space system over one RF period yields two dominant motional branches with secular-frequency estimates of 1.8137 and 1.8181 MHz. A third, predominantly axial branch exhibits a nominal frequency near 7.90 kHz, although its quantitative resolution remains limited by the residual Hessian trace and phase-return drift. Furthermore, a four-source electrostatic model is developed to investigate the influence of prescribed localized surface potentials on modal electric-field responses. Parametric analysis reveals correlation-dependent suppression and enhancement of normalized response coefficients, with multipliers ranging from approximately 0.5 to 1.5. This work establishes a computational framework for investigating nanoscale inclined blade-electrode architectures and their motional and surface-potential responses, while identifying important numerical convergence requirements for reliable three-dimensional trap characterization.

## 1. Introduction

Radio-frequency (RF) ion traps confine charged particles by combining time-dependent electric fields with the dynamical stability properties of their equations of motion. The resulting secular motion is central to trapped-ion spectroscopy, quantum information processing, and precision measurement [1-6]. In a practical trap, electrode layout determines the spatial RF field, the local confinement directions, the secular frequencies, and the potential barriers that delimit the region of stable motion. Alternative electrode geometries should therefore be evaluated by their actual three-dimensional fields rather than only by an idealized quadrupole model.

The scaling of trapped-ion systems motivates designs that provide useful confinement in compact geometries while retaining controllable motional modes. Numerical modelling is an important design step because it allows the electric-field structure and local curvature to be examined before fabrication. However, a local pseudopotential minimum or a one-dimensional barrier scan does not, by itself, demonstrate a global three-dimensional trapping depth; a lowest escape saddle may lie away from the selected path. Similarly, the small-q secular-frequency approximation is not a substitute for solving the RF-periodic equations when the relevant Mathieu parameters are not asymptotically small [1,2,7-9].

This study examines a proposed four-electrode architecture in which they are inclined relative to the reference geometry. The quantitative RF analysis is restricted to $\theta = 6°$, $d = 20$ μm, $V_{RF} = 150$ V, and $f_{RF} = 15$ MHz. The study has three primary aims: (i) calculate the local RF pseudopotential and estimate barriers along two in-plane diagonal paths; (ii) evaluate the nominal motional stability using the full symmetric RF-potential Hessian and a Floquet-period calculation of the associated matrix Mathieu equation; and (iii) assess how a prescribed four-source localized surface-potential model responds to selected assumptions about source correlations and spectral shape.

Analytical studies of surface-electrode ion traps have developed electrostatic models for relating electrode geometry to ion position, trap depth, and stability, while experimental work has demonstrated planar and microfabricated architectures [10-15]. Optimization studies and fabrication-oriented investigations have extended these approaches to electrode lattices, ion transport, design guidelines, and integrated trap technologies [16-20].

The electric-field noise near ion-trap surfaces is experimentally important, but its microscopic origin and spatial statistics are not universal; measured noise can depend on surface preparation, frequency, temperature, electrode distance, and multiple competing mechanisms [21-32]. Accordingly, the localized-potential calculation below is treated as a normalized response and sensitivity study. It does not assume that a four-patch model constitutes a complete physical model of electrode noise, and it does not predict absolute heating without an experimentally measured noise spectrum.

In this study, we employ a computational simulation framework to investigate RF null and pseudopotential landscape, Mathieu stability, Floquet analysis, and surface-potential response. The results should therefore be read as a numerical assessment of one proposed geometry and operating point, with explicit attention to the remaining uncertainty in the weak mode and to the sensitivity of the electrostatic response to computational-domain size.

The electrode arrangement and a conceptual illustration of the opposing RF phases are shown in Figs. 1 and 2, respectively.

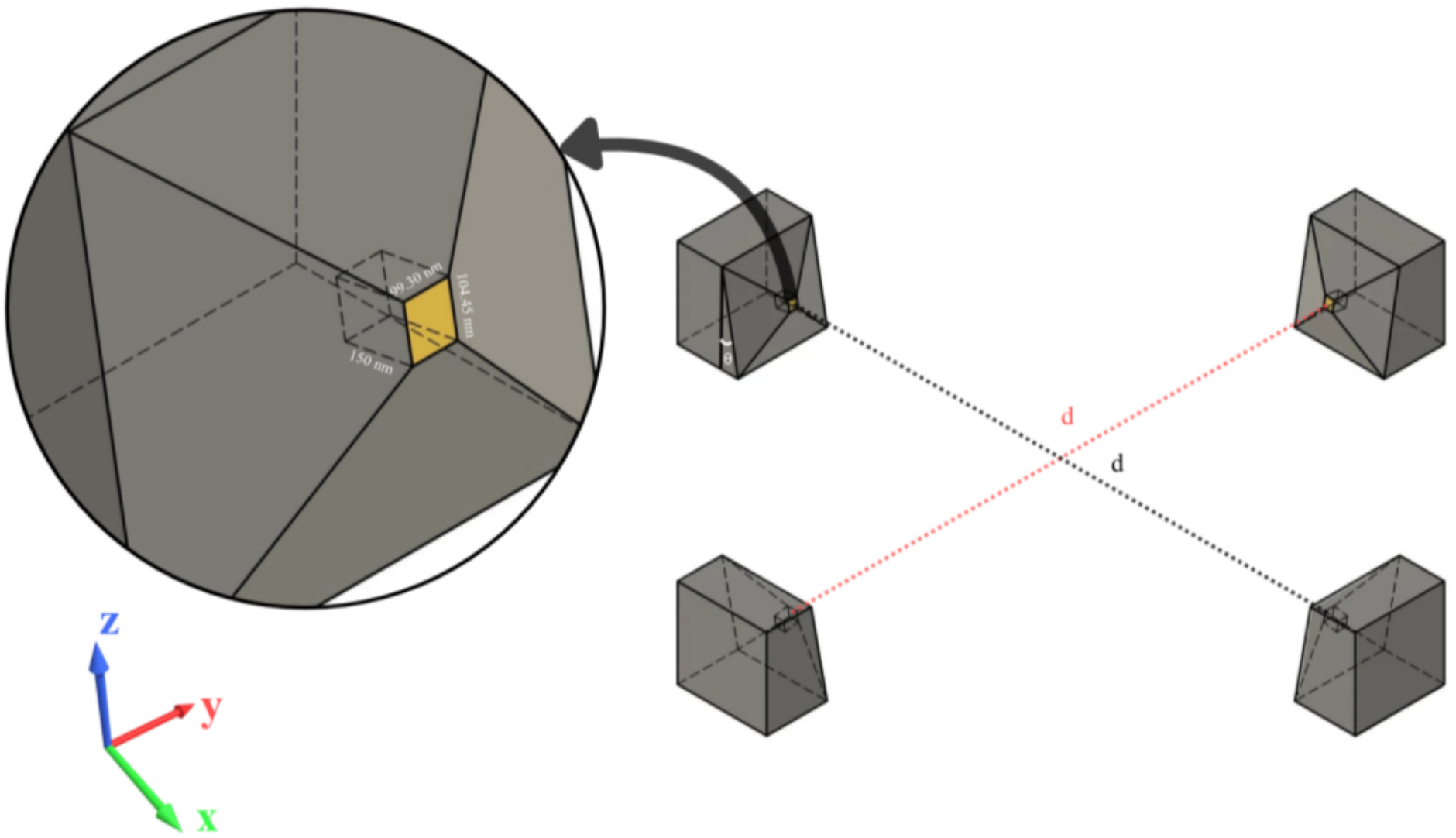

**Fig. 1.** Geometry of the proposed inclined four-blade electrode architecture. The angle θ denotes the electrode inclination and d denotes the full opposing-electrode separation as defined in the CAD model. The inset enlarges the exposed vacuum-facing face used for the localized surface-potential response calculation; its measured bounding extents are approximately 99.30, 104.45, and 150 nm along the two indicated directions.

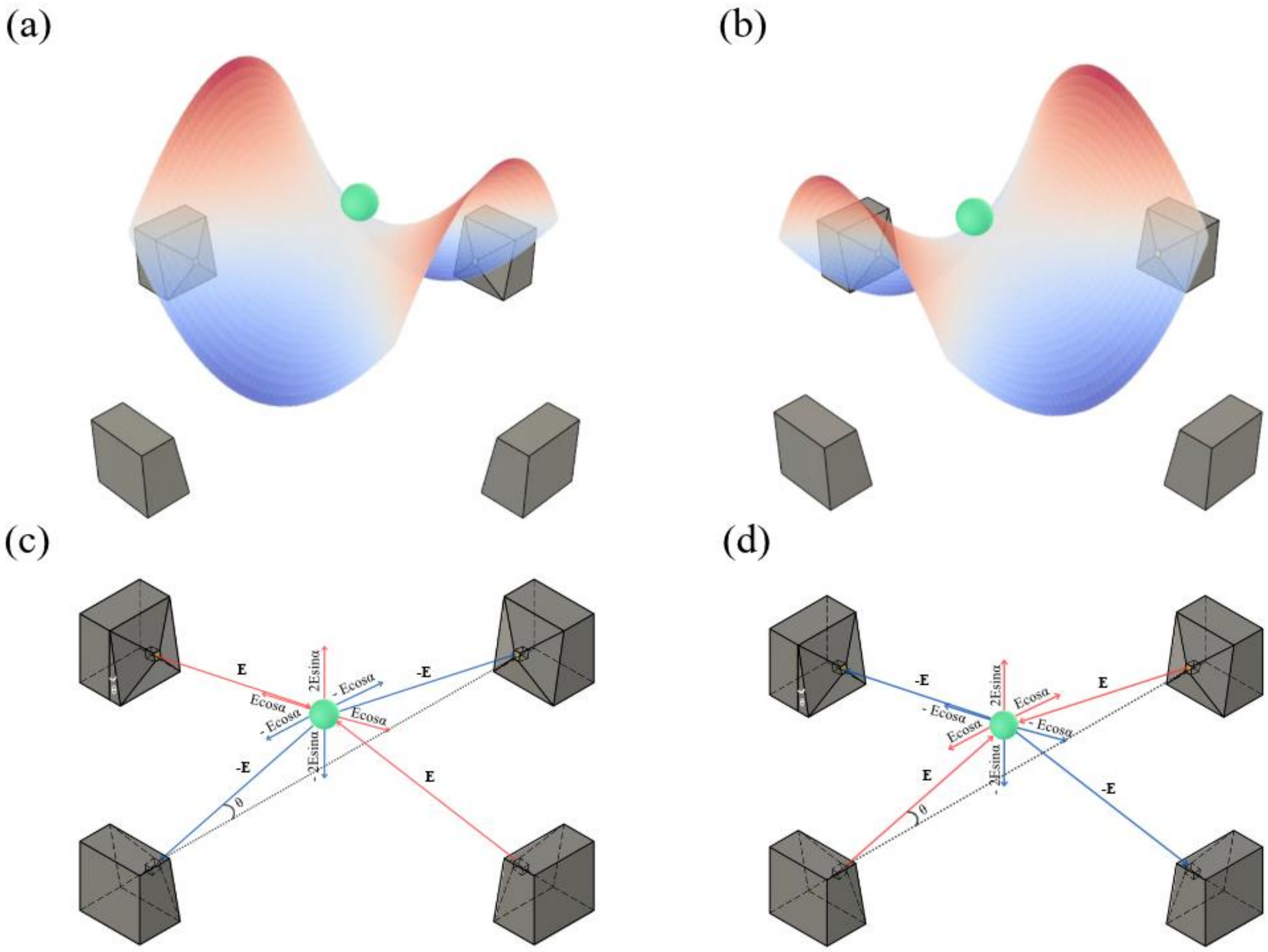

**Fig. 2.** Conceptual illustration of the opposing RF phases, qualitative potential landscape, and electric-field direction reversal. The colored surfaces and arrows are schematic, not numerical COMSOL results, where red and blue indicate higher and lower potentials, respectively. Quantitative RF-field, pseudopotential, Floquet, and noise-sensitivity results are presented separately in Figures 3-8.

## 2. Methods

### 2.1. Geometry and finite-element model

The three-dimensional electrode geometry was constructed in Autodesk Fusion 360 and imported as a STEP model into COMSOL Multiphysics 6.0 [33,34]. Numerical modelling is widely used to assess ion-trap electrode layouts before fabrication; related surface-trap design work has also used boundary-element approaches [19], the present fields were calculated using the finite-element model described here. The present RF analysis uses $\theta = 6°$ and $d = 20$ µm, where d denotes the full opposing-electrode separation as defined in the imported CAD geometry. The electrodes were assigned gold, and the surrounding domain was treated as vacuum. The selected ion was $^{171}Yb^{+}$, with charge +e and mass $m = 2.8395 \times 10^{-25}$ kg. The RF amplitude was 150 V (peak amplitude) and $f_{RF} = 15$ MHz, corresponding to $\Omega = 2\pi f_{RF} = 9.4248 \times 10^{7}$ rad s$^{-1}$.

**Table I.** Selected operating point and principal simulation inputs.

| Parameter | Value | Note |
|---|---|---|
| Ion | $^{171}Yb^{+}$; $m = 2.8395 \times 10^{-25}$ kg | Single positive charge |
| Geometry | $\theta = 6°$; $d = 20$ µm | Full opposing-electrode separation as defined in the CAD geometry |
| RF drive | $V_{RF} = 150$ V peak; $f_{RF} = 15$ MHz | $\Omega = 9.4248 \times 10^{7}$ rad s$^{-1}$; opposing pairs use opposite phase |

For the localized surface-potential response calculation, the surrounding computational enclosure was represented initially by a block of width 35 µm, depth 35 µm, and height 40 µm centred at (0, 0, 1 µm). The outer boundary was grounded in that response problem as a numerical truncation assumption; it is not asserted to represent a physically grounded chamber wall. A separate response calculation enlarged the block to 45 x 45 x 50 µm$^3$ while keeping its center and nominal local mesh-refinement settings unchanged. The effect is reported as a domain-size sensitivity test [35].

## 2.2. RF field and pseudopotential

Opposing electrode pairs were driven with opposite RF phase. At the reference phase, one pair was assigned $+V_{RF}$ and the other $-V_{RF}$. The time-dependent electric-potential distribution was calculated using the Electric Currents (ec) physics interface in COMSOL Multiphysics 6.0 with a Time Dependent study. The RF electrode boundary potentials were prescribed with opposing cosine waveforms. The time-dependent electric field was obtained from the solved potential as $E(r,t) = -\nabla V(r,t)$. For the applied cosine drive, $\Omega t = 0$ is the positive peak phase of the waveform. The electric-field and potential profiles at this phase are denoted by $E_{RF}(r)$ and $\Phi_{RF}(r)$, respectively; their peak-phase relation is given below.

$$E_{RF}(\mathbf{r}) = -\nabla\Phi_{RF}(\mathbf{r}).$$

The electric-field components and the spatial derivatives of the electric potential were evaluated near the calculated RF null. The six independent entries of the symmetric RF-potential Hessian were extracted from the time-dependent solution at the peak-drive reference phase $\Omega t = 0$ and used to construct the matrix Mathieu parameters. Results at the other sampled phases were retained to check the expected RF sign reversal and phase return. The Electric Currents (ec) interface was solved directly in the time domain. The solution was sampled over one RF period at $\Omega t = 0$, $\pi/2$, $\pi$, $3\pi/2$, and $2\pi$. For the sinusoidal excitation, the $\Omega t = 0$ field was used as the peak RF-field amplitude in the pseudopotential calculation, while the phase-resolved potential derivatives were used to assess the temporal consistency of the RF-potential Hessian.

For a singly charged $^{171}Yb^{+}$ ion, the effective RF pseudopotential was calculated from the peak RF electric-field amplitude extracted from the time-dependent Electric Currents solution at $\Omega t = 0$ [1,2,7]. In COMSOL, the field components used were ec.Ex, ec.Ey, and ec.Ez evaluated at this peak-drive phase:

$$\Psi(r) = e^2 |E_{RF}(r)|^2 / (4\, m\, \Omega^2).$$

This expression uses the peak amplitude of the RF field. The calculated pseudopotential was evaluated in the vacuum region around the RF-null estimate. The reported null was $r_0 = (-2.7172 \times 10^{-5}, -1.98631 \times 10^{-3}, 1.0734905)$ µm. The small x and y offsets are close to the coordinate origin relative to the micrometer-scale model. The potential minimum at this point was $\Psi_{min} = 4.2051 \times 10^{-26}$ J.

## 2.3. Directional pseudopotential barriers

To compare confinement along two in-plane directions, the pseudopotential was sampled along the straight paths $x = y$ and $x = -y$ in the xy plane. For each path, the directional barrier was taken as the lower of the first relevant side-peaks encountered while moving outwards from the central minimum. The barrier relative to the minimum is

$$\Delta U_{path} = \Psi_{max} - \Psi_{min}.$$

This procedure gives two path-specific estimates. The barrier along the z-axis was irrelevant because the pseudopotential across it was decreasing along the axis; see Fig. 4. Therefore, this estimation does not show a global three-dimensional saddle search.

## 2.4. Matrix Mathieu parameters and Floquet propagation

The local curvature of the RF potential was represented by the symmetric Hessian $H_{RF} = \nabla\nabla\Phi_{RF}(r_0)$, whose entries have units of $Vm^{-2}$. At the reference phase $\Omega t = 0$, the extracted matrix was

$$\begin{aligned} H_{RF} \approx [[&-2.630714 \times 10^{9}, 4.780776 \times 10^{5}, -5.302369 \times 10^{5}],\\ [&\ 4.780776 \times 10^{5}, 2.636738 \times 10^{9}, -8.756727 \times 10^{6}],\\ [&-5.302369 \times 10^{5}, -8.756727 \times 10^{6}, 1.175797 \times 10^{7}]]\ Vm^{-2}. \end{aligned}$$

Because the evaluation point lies in the charge-free vacuum, the exact electrostatic solution should satisfy Laplace's equation, $\nabla^2\Phi_{RF} = \mathrm{tr}(H_{RF}) = 0$. The trace of the numerically extracted Hessian was therefore retained as a diagnostic of the local field-solution consistency; it is particularly important when assessing the much weaker third curvature.

With $\tau = \Omega t/2$, the RF-only linearized equation was written in matrix Mathieu form as

$$d^2u/d\tau^2 + [A - 2\,Q\cos(2\tau)]\,u = 0, \quad A = 0,$$

where the dimensionless RF matrix is $Q = -2eH_{RF}/(m\Omega^2)$. All off-diagonal entries were retained. The resulting matrix was

$$\begin{aligned} Q \approx [[&0.334216, -6.074 \times 10^{-5}, 6.736 \times 10^{-5}],\\ [&-6.074 \times 10^{-5}, -0.334981, 1.112 \times 10^{-3}],\\ [&6.736 \times 10^{-5}, 1.112 \times 10^{-3}, -0.001494]]. \end{aligned}$$

The eigenvalues of Q define the principal Mathieu parameters in this RF-only formulation. The three nominal values are $q_{x\text{-like}} \approx +0.334216$, $q_{weak} \approx -0.001490$, and $q_{y\text{-like}} \approx -0.334985$. The strong eigenvectors are predominantly x- and y-directed; the weak eigenvector is predominantly z-directed. The mode labels here follow the eigenvectors of the displayed matrix rather than a frequency ordering.

The fundamental matrix of the periodic equation was propagated over one Mathieu period, $\Delta\tau = \pi$. Its Floquet multipliers $\mu$ are the eigenvalues of the monodromy matrix. In a stable oscillatory branch, the conjugate multiplier pair lies on the unit circle, $\mu = \exp(\pm i\beta\pi)$. The secular-frequency estimate is $f_{sec} = \beta f_{RF}/2$, with $\beta = |\arg\mu|/\pi$. This phase relation was used to obtain the finite-q frequencies. The mathematical stability test follows the standard Mathieu and Floquet framework [1,7-9].

As a separate numerical consistency check, the dominant Hessian derivatives recorded at $\Omega t = 0$ and $2\pi$ were compared. The observed phase-return drift is treated as a numerical limitation, not as validation of the weak mode.

## 2.5. Prescribed localized surface-potential response model

A separate electrostatic calculation was used to quantify the linear response to four prescribed localized surface-potential profiles on the exposed vacuum-facing faces associated with the -Y, +Y, -X, and +X electrode orientations. The profile for source j was

$$V_j(r) = V_{amp}\exp[-|r - r_j|^2/(2\ell^2)],$$

with $V_{amp} = 1$ V and reference width $\ell = 1.5$ nm. The centre coordinates were (0, −9.994540568, 1.021773398) μm for -Y; (0, +9.994540568, 1.021773398) μm for +Y; (−9.994540568, 0, 1.021773398) μm for -X; and (+9.994540568, 0, 1.021773398) μm for +X. The exposed -Y face used in the geometry check had area $1.0373 \times 10^{-14}$ $m^2$ and approximate x- and z-bounding extents 0.0993 and 0.1039 μm. The Gaussian integral at $\ell = 1.5$ nm was $1.4137 \times 10^{-17}$ $m^2$, close to $2\pi\ell^2$.

The 1 V amplitude is a normalization for computing linear response; it is not a measured patch-potential amplitude or voltage-noise PSD. One source was activated at a time, while the other relevant electrode boundaries were grounded. The spatially varying Dirichlet profile is interpreted as an effective localized surface-potential perturbation, not as a literal voltage discontinuity across part of a continuous ideal conductor. A physical interpretation would require a model of the origin and spatial statistics of these effective surface potentials; patch-potential, adsorbate, finite-electrode, and dielectric-layer studies illustrate mechanisms that may need to be distinguished [36-40].

The reference four-source response vectors were evaluated using a nominal local maximum element-size target of 0.35 nm near the active Gaussian-profile centre. The field-response vector at the ion was defined as $g_j = E_j(r_0)/V_{amp}$, with components in $m^{-1}$. Projections onto two approximate, real, approximately unit-normalized in-plane mode vectors were calculated as $c_{ij} = u_i \cdot g_j$. The vectors were $u_x = (0.999999976, -9.0433 \times 10^{-5}, 2.0035 \times 10^{-4})$ and $u_y = (9.1100 \times 10^{-5}, 0.999994437, -3.3344 \times 10^{-3})$. These are approximate coordinate-aligned projections derived from the curvature representation, not fully converged Floquet mode functions.

For independent sources with equal normalized spectra, the geometric factor is $K_{i,geom} = \Sigma_j |c_{ij}|^2$. For correlated sources, with $S_{v,jk}(f) = \langle V_j(f)V^*_k(f)\rangle$, the projected one-sided field-noise PSD is

$$S_{E,i}(f) = \Sigma_j \Sigma_k c_{ij} c^*_{ik} S_{v,jk}(f).$$

The four source spectra were assigned a common power-law shape $S_v(f) = S_{v0,norm} (f/f_0)^{-\alpha}$, with $f_0 = 1$ MHz and $S_{v0,norm} = 1\ V^2\ Hz^{-1}$ used as an arbitrary reference normalization. Writing $S_{v,jk} = S_v(f)\rho_{jk}(f)$ yields $S_{E,i}(f) = S_v(f)K_i(f)$, where $K_i(f) = \Sigma_{jk} c_{ij} c^*_{ik}\rho_{jk}(f)$. In the present sensitivity sweep, ρ was taken to be real and frequency-independent. The correlation matrix was Hermitian, had unit diagonal entries, and was positive semidefinite for the tested block-diagonal pairwise cases.

$$K_{i,corr} = K_{i,geom} + 2\ \Sigma_{(j<k)}\ \rho_{jk}\ \mathrm{Re}(c_{ij}\ c^*_{ik}).$$

Because the current $c_{ij}$ values are real, this reduces to the real-valued expression used in the COMSOL parameters. Only the -Y/+Y and -X/+X pair correlations were varied; all other off-diagonal correlations were set to zero. The normalized correlation multiplier was $R_i = K_{i,corr}/K_{i,geom}$. The factorization into spectral and geometric terms relies on the common-spectrum and frequency-independent-correlation assumptions and is not a general model of electrode surface noise.

For a one-sided electric-field-noise PSD evaluated at secular frequency $f_i$, the nominal secular-frequency heating relation was used in the adopted convention [21-23]:

$$\dot{n}_{i,model} = [e^2/(4m\hbar\omega_i)]\ S_{E,i}(f_i), \qquad \omega_i = 2\pi f_i.$$

The reported coefficient $C_i = \dot{n}_{i,model}/S_{v0,norm}$ is a normalized secular-frequency heating coefficient. It is not an absolute heating prediction because $S_{v0,norm}$ was arbitrarily set to $1\ V^2\ Hz^{-1}$, the source statistics are prescribed, and the present relation does not include a full RF micromotion-sideband sum [21,31].

## 2.6. Axisymmetric planar-reference benchmark

An independent two-dimensional axisymmetric electrostatic model was constructed to check the response to a radially symmetric Gaussian boundary profile with $V_{amp} = 1$ V and $\ell = 1.5$ nm against the leading-order infinite-plane solution. The vacuum domain was a rectangle spanning $0 \le r \le 50$ μm and $0 \le z \le 50$ μm, with relative permittivity $\varepsilon_r = 1$. The symmetry axis $r = 0$ used the Axial Symmetry condition; the lower boundary $z = 0$ was assigned $V(r,0) = V_{amp} \exp[-r^2/(2\ell^2)]$, and the upper boundary $z = 50$ μm and outer radial boundary $r = 50$ μm were grounded. The stationary solution was evaluated on the axis at $r = 0$ and $z = h = 9.993$ μm. The triangular mesh contained 27,206 domain elements and 526 boundary elements, with 54,939 degrees of freedom; a point-local maximum element-size target of 0.35 nm was specified at the Gaussian centre. This finite-domain model is an implementation check, not a validation of the three-dimensional inclined-electrode response.

For $\ell \ll h$, the leading-order ideal-plane far-field estimates at the symmetry axis are $V(0,h)/V_{amp} \approx \ell^2/h^2$ and $E_z(0,h)/V_{amp} \approx 2\ell^2/h^3$. These expressions provide a reference for checking the axisymmetric implementation; the finite grounded boundaries are retained in the COMSOL model [39].

# 3. Results

## 3.1. RF null and pseudopotential landscape

For the selected operating point ($\theta = 6°$, d = 20 μm, $V_{RF}$ = 150 V, $f_{RF}$ = 15 MHz), minimization of the RF-field magnitude gave $r_0 = (-2.7172 \times 10^{-5}, -1.98631 \times 10^{-3}, 1.0734905)$ μm. The pseudopotential minimum was $4.2051 \times 10^{-26}$ J. The RF-null estimate lies close to the CAD reference point near (0, 0, 1 μm), relative to the overall micrometre-scale geometry.

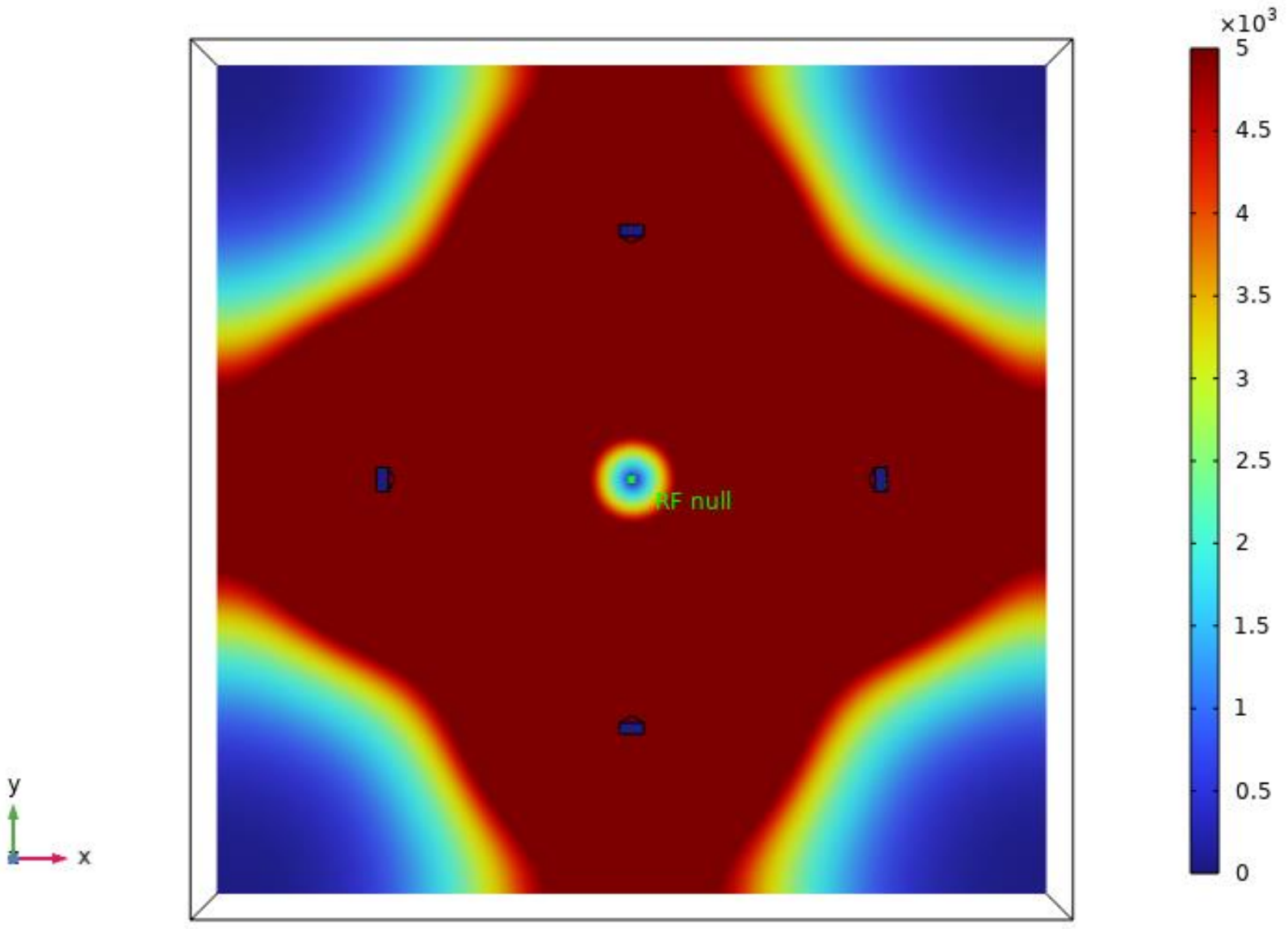

**Fig. 3.** Magnitude of the RF electric-field amplitude, $|E_{RF}|$ (V m⁻¹), around the trapping region (the view is of xy-plane) at $\theta = 6°$, d = 20 μm, $V_{RF}$ = 150 V, and $f_{RF}$ = 15 MHz, with the calculated RF-null position $r_0$ marked.

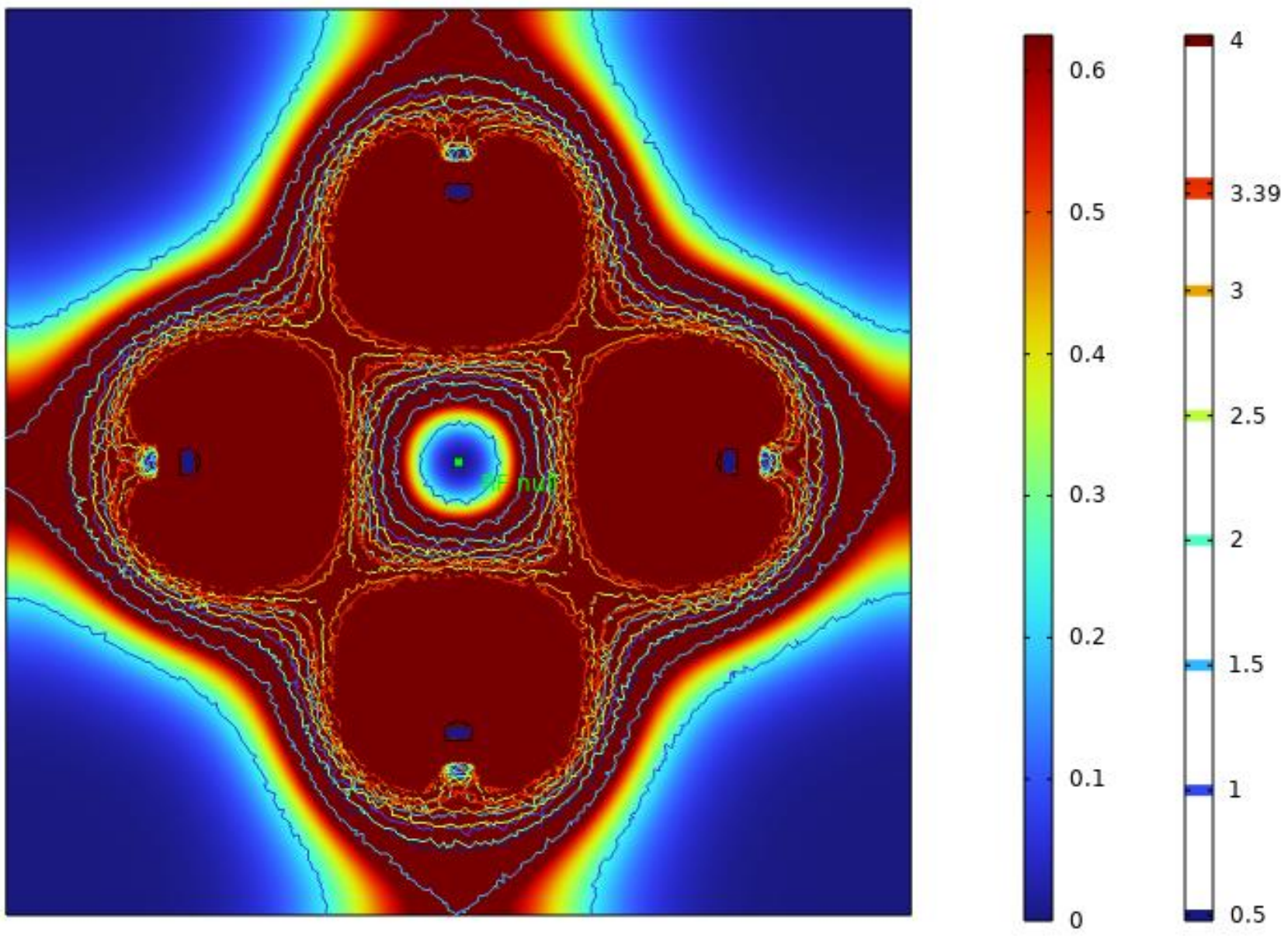

**Fig. 4.** Relative xy-plane pseudopotential, $U_{ps} - U_{min}$ (meV), around the RF null for the selected operating point. The figure uses the x and y axes and marks $r_0$.

## 3.2. Directional in-plane barrier estimate

The selected lower side-peaks along the two in-plane diagonals give directional barrier values summarized in Table II. The x = y path has $\Delta U = 5.49565 \times 10^{-22}$ J, corresponding to 3.430 meV. The x = -y path has $\Delta U = 5.43050 \times 10^{-22}$ J, corresponding to 3.389 meV. The two results differ by approximately 1.2%, indicating similar barrier scales along these two specified paths.

Fig. 5 compares the two one-dimensional profiles and identifies the lower side-peak used to define each directional barrier. The profiles shown in Fig. 5 were extracted from the corresponding COMSOL line evaluations.

**Table II.** Directional pseudopotential barrier estimates for the two in-plane paths.

| Path | $U_{min}$ (J) | Selected $U_{max}$ (J) | ΔU (J) | ΔU (meV) |
|---|---|---|---|---|
| x = y | 4.2051 x $10^{-26}$ | 5.49607 x $10^{-22}$ | 5.49565 x $10^{-22}$ | 3.430 |
| x = -y | 4.2051 x $10^{-26}$ | 5.43092 x $10^{-22}$ | 5.43050 x $10^{-22}$ | 3.389 |

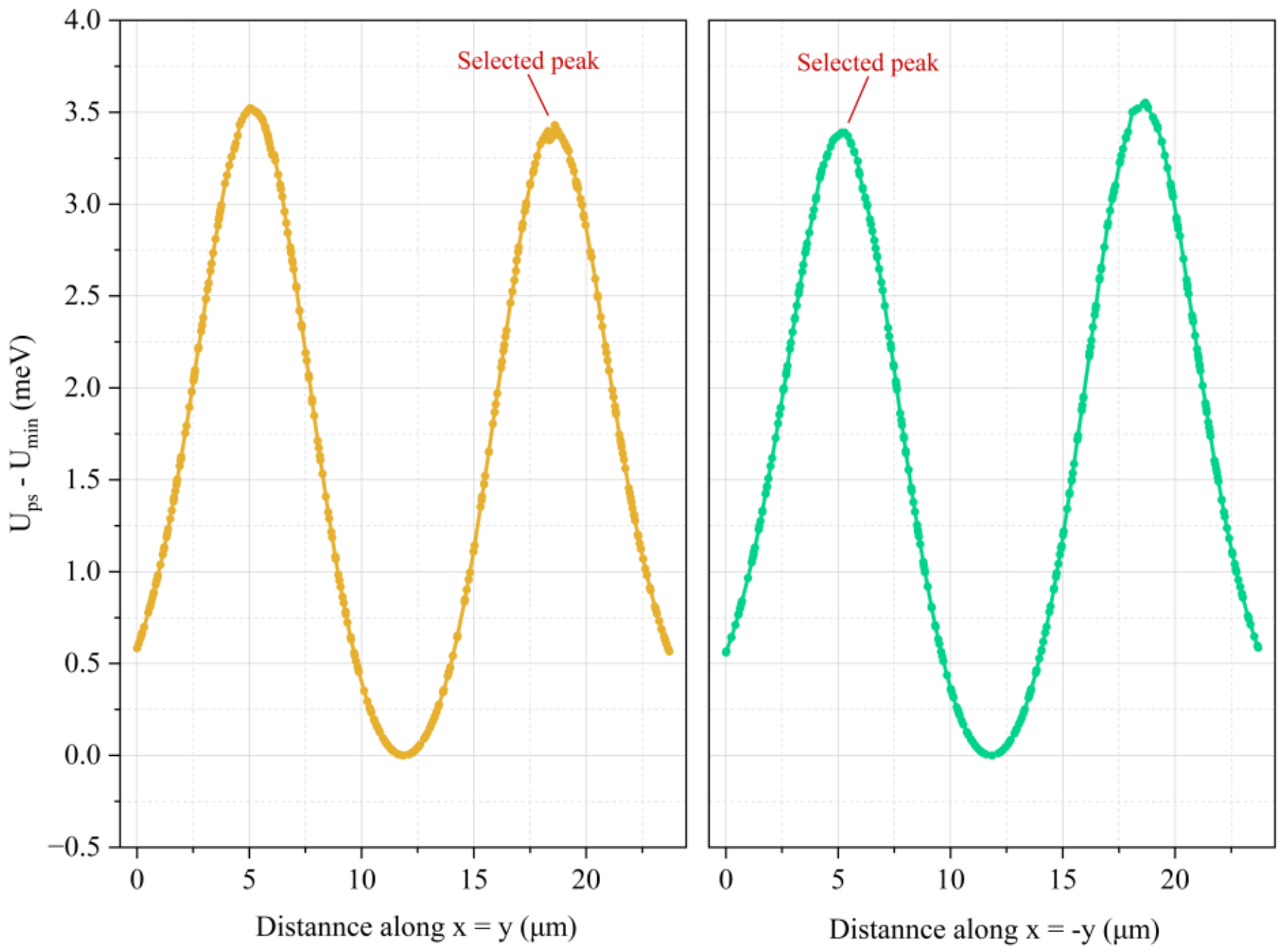

**Fig. 5.** RF pseudopotential profiles along two in-plane diagonal directions through the central trapping region for the proposed inclined four-blade trap at θ = 6°, d = 20 μm, $V_{RF}$ = 150 V and $f_{RF}$ = 15 MHz. Left one denotes the profile along the direction parallel to x = y. Right one denotes the profile along the direction parallel to x = -y. The pseudopotential is referenced to its value at the calculated RF null. The selected lower side-peaks correspond to directional barrier estimates of 3.430 and 3.389 meV, respectively.

## 3.3. Mathieu parameters and nominal Floquet stability

The extracted RF-potential Hessian has two dominant curvatures of approximately -2.631 x $10^9$ and +2.637 x $10^9$ V $m^{-2}$, while the weak third diagonal curvature is approximately 1.176 x $10^7$ V $m^{-2}$. With the full Hessian converted to the dimensionless matrix Q, its eigenvalues are approximately +0.334216, -0.001490, and -0.334985. Thus, the maximum absolute principal Mathieu parameter is $q_{max} \approx 0.334985$.

The nominal Floquet-period calculation produced complex-conjugate multiplier pairs with moduli approximately equal to unity for all three principal directions. The two strong branches correspond to secular-frequency estimates of 1.8137 and 1.8181 MHz. For comparison, the lowest-order RF-only small-|q| estimate, $f_{sec}^{(small\text{-}q)} \approx |q|\, f_{RF}/(2\sqrt{2})$, gives approximately 1.7724 and 1.7765 MHz for the two strong branches. The Floquet values are therefore about 2.3% higher, illustrating why the finite-q calculation is preferable at $|q| \approx 0.335$ rather than treating the small-q expression as exact.

The weak branch gives a nominal value near 7.90 kHz, but it is not quantitatively resolved. In the charge-free vacuum region, the Hessian trace residual is approximately 1.78 x $10^7$ V $m^{-2}$, about 1.51 times the weak curvature of approximately 1.18 x $10^7$ V $m^{-2}$ and about 0.67% of the largest diagonal curvature. The difference between the $\Omega t = 0$ and $2\pi$ dominant Hessian components is also about 0.5% of the dominant curvature, corresponding to a change of order 1.31 x $10^7$ V $m^{-2}$, or about 1.11 times the weak curvature. This phase-return drift is therefore also significant for the weak branch. Accordingly, the results support nominal oscillatory dynamics in the two dominant directions, whose frequency estimates are numerically more reliable than the weak-mode estimate.

**Table III.** Principal Mathieu parameters and nominal Floquet results.

| Principal branch | $q_i$ | Representative μ pair | \|μ\| | $f_{sec}$ estimate | Assessment |
|---|---|---|---|---|---|
| x-like strong | +0.334216 | 0.725026 ± 0.688722i | ≈ 1 | 1.8137 MHz | Nominally stable; better resolved |
| weak, mainly z | -0.001490 | 0.999995 ± 0.003310i | ≈ 1 | 7.90 kHz | Provisional; curvature not resolved |
| y-like strong | -0.334985 | 0.723762 ± 0.690050i | ≈ 1 | 1.8181 MHz | Nominally stable; better resolved |

The multipliers listed in Table III are approximate values from the recorded Floquet output and are rounded for presentation. Fig. 6 plots these representative conjugate pairs against the unit circle; the plotted coordinates should therefore be read at the precision of the tabulated multiplier values. The close-to-unit-modulus result is a property of the nominal extracted Mathieu model; it does not remove the numerical uncertainty associated with the weak curvature.

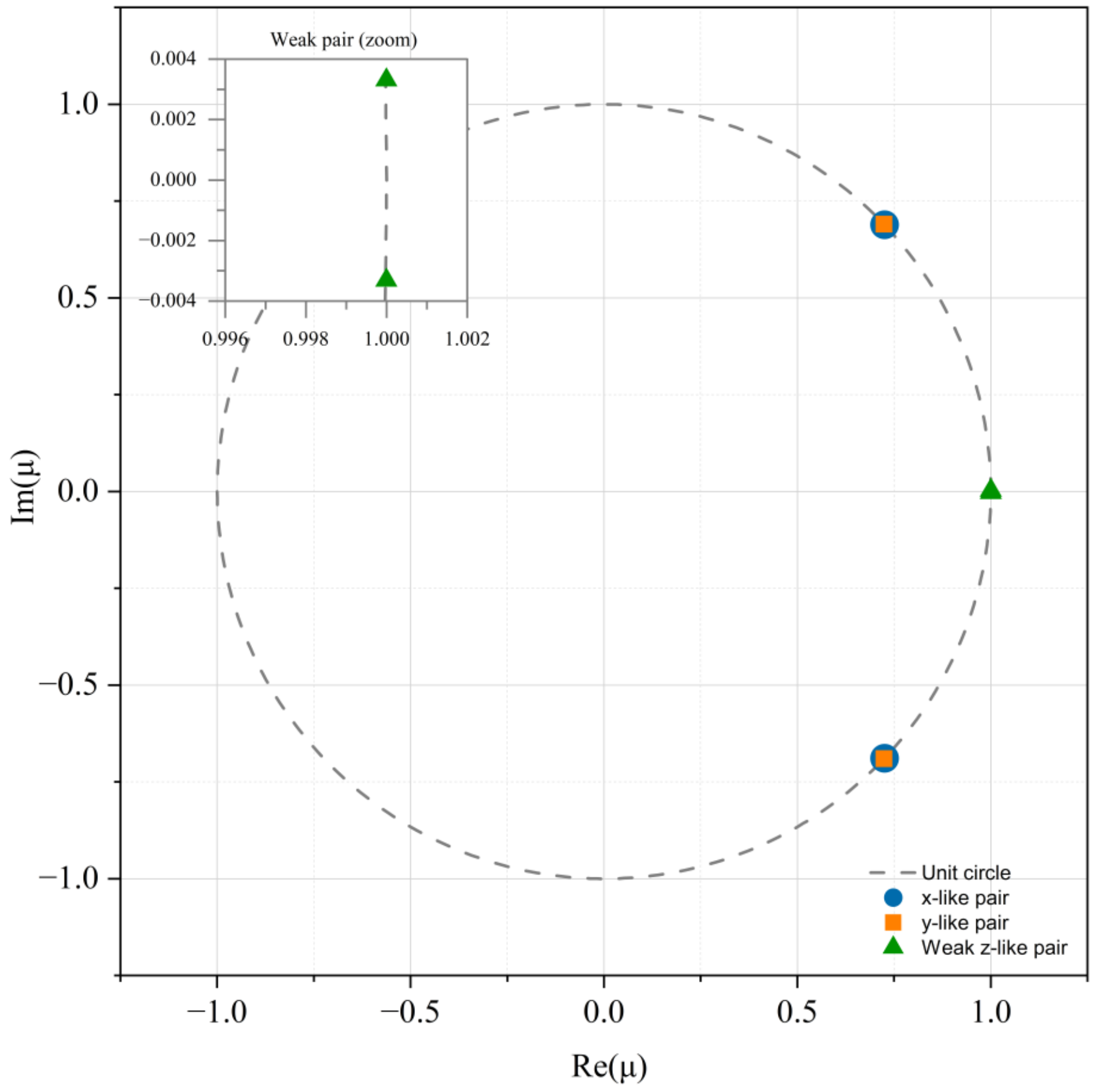

**Fig. 6.** plots the unrounded complex-conjugate Floquet multiplier pairs obtained directly from the numerical eigenvalue analysis of the monodromy matrix. The corresponding values in Table III are rounded to six decimal places for presentation. Although all three nominal multiplier pairs lie close to the unit circle, the weak z-like branch remains under-resolved because of the numerical uncertainty in its extracted curvature.

## 3.4. Normalized localized surface-potential response

### 3.4.1. Field-response vectors and geometric factors

The four prescribed Gaussian profiles were applied individually to their exposed electrode faces. Table IV reports the raw electric-field response per unit peak potential and the approximate modal projections at the ion. The x-like response is dominated by the two X-oriented sources, and the y-like response is dominated by the two Y-oriented sources. The small transverse components were retained in the calculations; their precision is lower than that of the dominant components.

**Table IV.** Raw normalized electric-field response vectors and approximate modal projections at ℓ = 1.5 nm.

| Source | $g_x$ ($m^{-1}$) | $g_y$ ($m^{-1}$) | $g_z$ ($m^{-1}$) | $c_{xj}$ ($m^{-1}$) | $c_{yj}$ ($m^{-1}$) |
|---|---|---|---|---|---|
| -Y | $-4.23 \times 10^{-7}$ | $+2.22425 \times 10^{-2}$ | $-1.16009 \times 10^{-4}$ | $-2.458 \times 10^{-6}$ | $+2.22427 \times 10^{-2}$ |
| +Y | $+1.332 \times 10^{-5}$ | $-2.22973 \times 10^{-2}$ | $-6.71319 \times 10^{-5}$ | $+1.532 \times 10^{-5}$ | $-2.22970 \times 10^{-2}$ |
| -X | $+2.22753 \times 10^{-2}$ | $-4.78719 \times 10^{-6}$ | $-7.47735 \times 10^{-5}$ | $+2.22753 \times 10^{-2}$ | $-2.509 \times 10^{-6}$ |
| +X | $-2.22758 \times 10^{-2}$ | $-3.38874 \times 10^{-5}$ | $-5.02581 \times 10^{-5}$ | $-2.22758 \times 10^{-2}$ | $-3.575 \times 10^{-5}$ |

For independent and equally normalized source spectra, the geometric factors calculated from the full-precision projection coefficients were $K_{x,geom} = 9.9240 \times 10^{-4}$ $m^{-2}$ and $K_{y,geom} = 9.9190 \times 10^{-4}$ $m^{-2}$. The near equality of these values reflects the approximately symmetric dominant response of the two opposing source pairs for the chosen geometry. These factors characterize only the specified boundary profiles and should not be interpreted as a universal surface-noise property of the trap.

### 3.4.2. Assumed spectral-exponent dependence

For the independent-source baseline, the normalized field-noise PSD was evaluated at $f_x = 1.814$ MHz and $f_y = 1.818$ MHz using α = 0.5, 1.0, and 1.5. Because both mode frequencies exceed the reference frequency $f_0 = 1$ MHz, the assumed power law decreases with increasing α. This is a direct consequence of the selected spectrum rather than a new measured property of the trap.

**Table V.** Normalized field-noise PSD and normalized secular-frequency heating coefficient for independent sources.

| α | $S_{E,x\ norm}$ ($V^2\ m^{-2}\ Hz^{-1}$) | $S_{E,y\ norm}$ ($V^2\ m^{-2}\ Hz^{-1}$) | $C_x$ [$s^{-1}/(V^2\ Hz^{-1})$] | $C_y$ [$s^{-1}/(V^2\ Hz^{-1})$] |
|---|---|---|---|---|
| 0.5 | $7.37 \times 10^{-4}$ | $7.36 \times 10^{-4}$ | $1.39 \times 10^{10}$ | $1.38 \times 10^{10}$ |
| 1.0 | $5.47 \times 10^{-4}$ | $5.46 \times 10^{-4}$ | $1.03 \times 10^{10}$ | $1.02 \times 10^{10}$ |
| 1.5 | $4.06 \times 10^{-4}$ | $4.05 \times 10^{-4}$ | $7.64 \times 10^{9}$ | $7.59 \times 10^{9}$ |

The coefficients $C_i$ are defined by $\dot{n}_{i,model} = C_i S_{V0,norm}$ under the stated spectrum and response assumptions. Since $S_{V0,norm} = 1$ $V^2\ Hz^{-1}$ is a unit reference rather than a measured source PSD, these values are normalized sensitivity coefficients, not absolute heating-rate predictions. A physical heating estimate would require a justified or measured PSD and a mode treatment consistent with the driven RF dynamics.

### 3.4.3. Correlation sensitivity

The correlation sweep varied only ρ_−Y,+Y and ρ_−X,+X. Other off-diagonal entries were zero. For each tested pair, ρ = 0 or ±0.5; the resulting two-by-two correlation blocks have eigenvalues 1 ± ρ and are positive definite for these values. The reported matrices are therefore valid correlation matrices for the selected block-diagonal scenarios.

**Table VI.** Correlation-weighted geometric factors and relative multipliers for the nine prescribed scenarios.

| Scenario | $\rho_{-Y,+Y}$ | $\rho_{-X,+X}$ | $K_{x,corr}$ ($m^{-2}$) | $K_{y,corr}$ ($m^{-2}$) | $R_x$ | $R_y$ |
|---|---|---|---|---|---|---|
| Independent | 0 | 0 | $9.924 \times 10^{-4}$ | $9.919 \times 10^{-4}$ | 1.0 | 1.0 |
| Y pair positive | +0.5 | 0 | $9.924 \times 10^{-4}$ | $4.9595 \times 10^{-4}$ | 1.0 | 0.5 |
| Y pair negative | -0.5 | 0 | $9.924 \times 10^{-4}$ | $1.4878 \times 10^{-3}$ | 1.0 | 1.5 |
| X pair positive | 0 | +0.5 | $4.962 \times 10^{-4}$ | $9.919 \times 10^{-4}$ | 0.5 | 1.0 |
| X pair negative | 0 | -0.5 | $1.4886 \times 10^{-3}$ | $9.919 \times 10^{-4}$ | 1.5 | 1.0 |
| Both positive | +0.5 | +0.5 | $4.962 \times 10^{-4}$ | $4.9595 \times 10^{-4}$ | 0.5 | 0.5 |
| Both negative | -0.5 | -0.5 | $1.4886 \times 10^{-3}$ | $1.4878 \times 10^{-3}$ | 1.5 | 1.5 |
| Mixed A | +0.5 | -0.5 | $1.4886 \times 10^{-3}$ | $4.9595 \times 10^{-4}$ | 1.5 | 0.5 |
| Mixed B | -0.5 | +0.5 | $4.962 \times 10^{-4}$ | $1.4878 \times 10^{-3}$ | 0.5 | 1.5 |

The multipliers are close to 0.5, 1.0, and 1.5 because the dominant opposing-pair projections have nearly equal magnitudes and opposite signs. Positive correlation between two oppositely signed dominant projections suppresses their contribution in the tested model, whereas negative correlation enhances it. The effect is modal because the X pair dominates the x-like projection while the Y pair dominates the y-like projection. The factors are conditional on the chosen four-source basis, common spectrum, real frequency-independent correlations, and approximate mode vectors; they are not evidence that the physical electrodes possess these correlations.

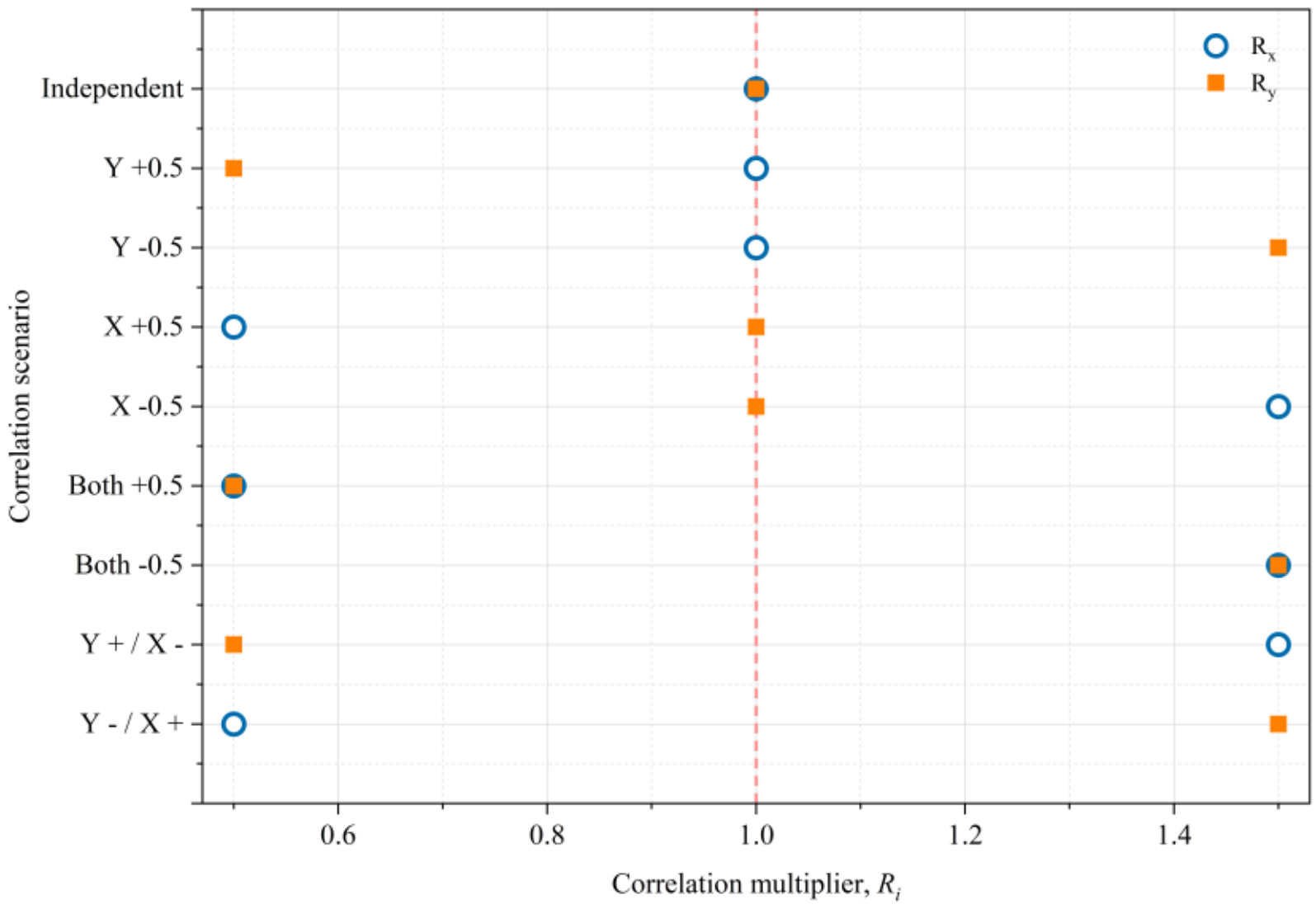

**Fig. 7.** Correlation multipliers $R_x$ and $R_y$ for the nine prescribed real pairwise-correlation scenarios considered in Table VI. The vertical dashed line represents the independent-source reference value, $R_i = 1$. Values below and above unity indicate suppression and enhancement, respectively, relative to the independent-source case. The results represent a model-sensitivity analysis based on prescribed electrode-pair correlations and do not correspond to experimentally measured noise correlations.

### 3.4.4. Mode-orientation and Gaussian-width sensitivity

For each in-plane rotation $\delta = \pm 0.1°$ and $\pm 1°$, the modal projections were recalculated as $c'_{xj} = \cos(\delta)c_{xj} + \sin(\delta)c_{yj}$ and $c'_{yj} = -\sin(\delta)c_{xj} + \cos(\delta)c_{yj}$. The small cross-projections changed appreciably relative to their original values and could change sign, but the dominant projections remained associated with the corresponding opposing source pair. Across the nine prescribed correlation scenarios, the maximum absolute change in either multiplier was approximately 5.7 x $10^{-6}$ for rotations of ±0.1° and 3.31 x $10^{-4}$ for rotations of ±1°. This is robustness to the tested algebraic rotations, not an uncertainty estimate for the actual Floquet modes; that estimate requires converged mode functions from the RF-periodic dynamics.

Gaussian-width sensitivity was tested for one Y-oriented patch and one X-oriented patch at $\ell$ = 1.0, 1.5, and 2.0 nm. The peak amplitude remained 1 V, while the geometry, selected face, ion evaluation point, and nominal local mesh-size target were held fixed. Since the surface integral of a localized two-dimensional Gaussian scales approximately as $\ell^2$, the area-rescaled diagnostic $|g^*_i(\ell)| = |g_i(\ell)|(1.5\ \text{nm}/\ell)^2$ was used to compare the dominant components at approximately equal integrated profile strength. Table VII lists the raw and rescaled responses.

**Table VII.** Gaussian-width dependence of the dominant responses and area-rescaled diagnostic.

| Patch / component | $\ell$ (nm) | Raw \|g\| ($m^{-1}$) | Area-rescaled \|g*\| ($m^{-1}$) | Difference from 1.5 nm |
|---|---|---|---|---|
| -Y, \|g_y\| | 1.0 | 9.779 x $10^{-3}$ | 2.200 x $10^{-2}$ | −1.08% |
| -Y, \|g_y\| | 1.5 | 2.224 x $10^{-2}$ | 2.224 x $10^{-2}$ | Reference |
| -Y, \|g_y\| | 2.0 | 3.955 x $10^{-2}$ | 2.225 x $10^{-2}$ | +0.02% |
| -X, \|g_x\| | 1.0 | 9.816 x $10^{-3}$ | 2.209 x $10^{-2}$ | −0.85% |
| -X, \|g_x\| | 1.5 | 2.228 x $10^{-2}$ | 2.228 x $10^{-2}$ | Reference |
| -X, \|g_x\| | 2.0 | 3.961 x $10^{-2}$ | 2.228 x $10^{-2}$ | +0.03% |

At fixed peak voltage, the raw dominant response rises substantially with ℓ. After area rescaling, the dominant components vary by less than about 1.1% over the three tested widths. This suggests that much of the raw dependence reflects the increasing integrated strength of the imposed profile. Area rescaling is only a diagnostic normalization; it does not physically justify the chosen Gaussian width or identify it with a measured surface-noise correlation length.

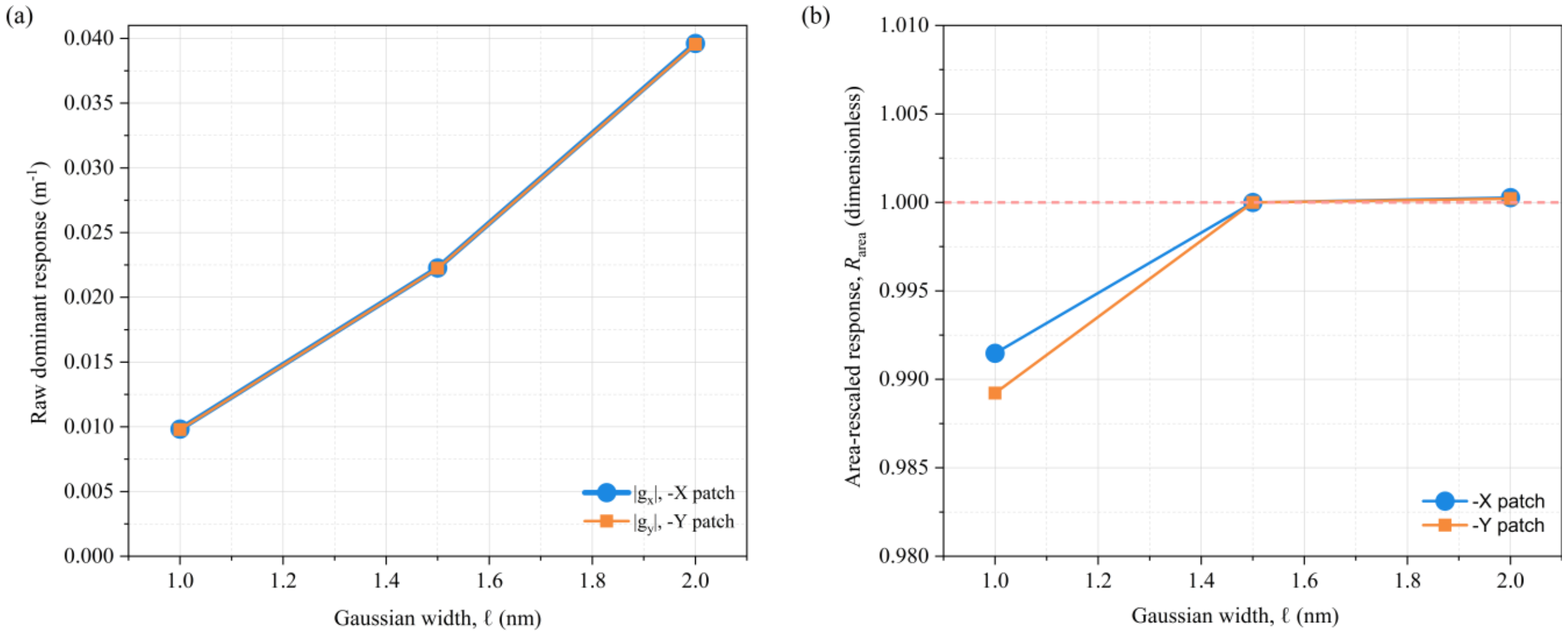

**Fig. 8.** Gaussian-width sensitivity of the dominant -X and -Y patch-response components at fixed peak surface-potential amplitude. (a) Raw dominant electric-field responses as a function of Gaussian width. (b) Area-rescaled responses, normalized independently to their respective values at a Gaussian width of 1.5 nm. The near-unity normalized responses indicate that the width dependence of the raw response is approximately consistent with the prescribed Gaussian-area scaling. These results characterize the prescribed boundary perturbation model and should not be interpreted as validation of a physical electrode-noise distribution.

A two-setting mesh comparison was available for the dominant -Y component at nominal local maximum element sizes of 0.50 and 0.35 nm. The changes in $|g_y|$ for ℓ = 1.0, 1.5, and 2.0 nm were approximately 0.42%, 0.62%, and 0.19%, respectively. These results support limited numerical stability of that dominant component between the two tested settings, but they do not establish formal convergence of the complete response vectors. The same two-level mesh comparison has not been completed for the -X response vector.

### 3.4.5. Computational-domain and planar-reference sensitivity

The effect of the grounded outer boundary was tested by increasing the computational block from 35 x 35 x 40 $\mu m^3$ to 45 x 45 x 50 $\mu m^3$ while keeping the block centre at (0, 0, 1 μm), the Gaussian boundary profile, and the ion evaluation point unchanged. Table VIII compares the normalized response. The potential at the ion increased by approximately 15.3%, and the dominant y-component increased by approximately 5.3%. The z-component magnitude decreased by approximately 25.0%. The relative change in $g_x$ is not meaningful as a percentage because its original value was close to zero; its magnitude remains much smaller than the dominant response.

**Table VIII.** Computational-domain sensitivity of the −Y Gaussian response.

| Quantity | 35 x 35 x 40 $\mu m^3$ | 45 x 45 x 50 $\mu m^3$ | Observed change |
|---|---|---|---|
| $V_2/V_{\text{amp at ion}}$ | $1.1864 \times 10^{-7}$ | $1.3680 \times 10^{-7}$ | +15.3% |
| $g_x$ (m⁻¹) | $-4.23 \times 10^{-7}$ | $-2.04 \times 10^{-5}$ | Both small vs $g_y$; relative change uninformative |
| $g_y$ (m⁻¹) | $2.2242 \times 10^{-2}$ | $2.3421 \times 10^{-2}$ | +5.30% |
| $g_z$ (m⁻¹) | $-1.1601 \times 10^{-4}$ | $-8.7057 \times 10^{-5}$ | 25.0% lower magnitude |

Only two domain sizes were examined, so this result demonstrates finite-domain sensitivity rather than convergence. The same four-patch response at the larger domain was not independently completed for every source, and the correlation table therefore uses the original-domain projections. This distinction should be preserved when interpreting the tables.

An independent 2D axisymmetric reference calculation was performed for a radially symmetric Gaussian boundary profile with $\ell = 1.5$ nm. At a height $h = 9.993$ μm, COMSOL gave $V/V_{amp} = 2.2403 \times 10^{-8}$ and $E_z/V_{amp} = 4.5139 \times 10^{-3}\ m^{-1}$.

The far-field ideal-planar estimates are approximately $\ell^2/h^2 = 2.2532 \times 10^{-8}$ and $2\ell^2/h^3 = 4.5095 \times 10^{-3}\ m^{-1}$; the axisymmetric values differ by about 0.6% for potential and 0.1% for field. This comparison checks the 2D reference implementation, but it does not validate the 3D inclined-electrode response.

In the original 3D model, the -Y Gaussian response was $V_2/V_{amp} \approx 1.1864 \times 10^{-7}$ and $g_y \approx 2.2242 \times 10^{-2}\ m^{-1}$, approximately 5.3 and 4.9 times the corresponding ideal-planar reference values. A separate uniform-potential excitation over the same exposed face was also evaluated. Its amplitude was rescaled using the measured face area $A_{42} \approx 1.0373 \times 10^{-14}\ m^2$ and the prescribed Gaussian integral $2\pi\ell^2$, giving a uniform-face amplitude of approximately $1.363 \times 10^{-3}$ V relative to the 1 V Gaussian reference. With this area-based rescaling, its normalized potential and dominant field component at the ion were about 13.0% and 14.6% higher than the Gaussian results. This comparison shows that the calculated response depends on the spatial distribution of the prescribed boundary excitation, but it does not explain the remaining factor-of-five difference relative to the ideal-plane estimate. That discrepancy remains unresolved and is not attributed solely to the outer boundary or to finite geometry.

# 4. Discussion

The present calculation characterizes one selected configuration of the proposed inclined four-blade architecture. The two in-plane diagonal scans yield barriers of similar magnitude, with the lower evaluated path barrier at 3.389 meV. These values are useful directional estimates for comparing paths within the calculated plane, but they do not establish a global three-dimensional escape barrier. The quantitative field and pseudopotential plots should therefore be examined alongside the diagonal line profiles before conclusions are drawn about the topology of the full trapping region.

The matrix Mathieu analysis is the principal dynamical result. The dimensionless Q matrix has two dominant eigenvalues near ±0.335 and a third nominal eigenvalue near −0.00149. The associated Floquet multipliers have nominal moduli close to unity, and the two dominant branches yield secular-frequency estimates near 1.814 and 1.818 MHz. These finite-q estimates are preferable to relying solely on a small-q formula at $|q| \approx 0.335$. However, a nonzero Hessian trace in the charge-free vacuum domain and the observed phase-return derivative drift are comparable to the weak-mode curvature. The weak-mode frequency and a definitive full three-dimensional stability statement therefore remain provisional. A higher-accuracy field solution with improved Laplace consistency and a converged full Hessian is required to resolve that direction.

The four-source model shows how source correlations can suppress or enhance the normalized field-noise response of a projected mode. In the prescribed cases, the multipliers range from approximately 0.5 to 1.5. The algebraic mode-rotation test shows that these factors are robust to the tested small in-plane rotations even though the minor cross-projections are individually sensitive. This robustness is conditional on the chosen source pairings and common-spectrum model. It should not be generalized to arbitrary correlated surface noise, which is more naturally described by a spatially resolved cross-spectral density $S_{V(r,r',f)}$ [21,36-40].

The width-sensitivity results show near-constant dominant responses after area-based rescaling, but this normalization does not make the Gaussian profile a physically validated surface-potential model. The selected width and the four discrete sources were prescribed for computational sensitivity analysis. No surface correlation length, noise PSD, or source-correlation coefficient was measured. Consequently, the normalized coefficients should not be presented as absolute heating-rate predictions. To obtain physical predictions, the model would need a justified surface-statistics model or measured spectra, a more complete treatment of RF-periodic motional modes, and a numerical response that passes domain and mesh convergence checks [35,39,40].

The domain-size test is particularly important: increasing the enclosure changes the calculated potential and dominant field response by measurable amounts. Although the dominant field change between the two 3D domains is smaller than the factor-of-five planar-reference discrepancy, the comparison does not yet isolate the cause of that discrepancy. It would be premature to attribute it to finite electrode geometry, boundary truncation, or normalization alone. The uniform-face comparison also shows that the spatial shape of the source matters even when the integrated boundary-potential strength is matched.

Taken together, the results support a cautious conclusion: the proposed architecture has a numerically characterized RF pseudopotential and two dominant motional branches with nominal Floquet multiplier pairs close to unit modulus and more reliable frequency estimates than the weak branch at the selected operating point, while the weak branch and the global trap depth remain unresolved. The four-patch calculation adds a conditional sensitivity analysis of prescribed localized surface-potential perturbations, but further physical and numerical validation is needed before it can be treated as a predictive surface-noise model.

# 5. Conclusions

A finite-element assessment of a proposed inclined four-blade ion trap was carried out at θ = 6°, d = 20 μm, $V_{RF}$ = 150 V, and $f_{RF}$ = 15 MHz for $^{171}Yb^{+}$. The RF null was found near (-2.72 x $10^{-5}$, -1.986 x $10^{-3}$, 1.07349) μm, and the RF pseudopotential minimum was approximately 4.2051 x $10^{-26}$ J. Straight-line scans along x = y and x = -y gave directional barrier estimates of 3.430 and 3.389 meV. These are the lowest barriers among the two paths examined, not a validated global three-dimensional trap depth.

The extracted RF Hessian produced principal Mathieu parameters of approximately +0.334216, -0.001490, and -0.334985. The nominal Floquet calculation yielded two strong secular-frequency estimates near 1.814 and 1.818 MHz, with multiplier pairs close to unit modulus. The weak mode near 7.9 kHz is not quantitatively resolved because the numerical curvature residual and phase-return drift are comparable to its curvature. Thus, the two strong branches are supported by the present model, but definitive three-dimensional stability requires a more converged electrostatic solution.

The four-source surface-potential calculation quantified how prescribed real pairwise correlations, assumed spectral exponents, approximate mode orientation, and Gaussian width affect normalized response coefficients. Its results are explicitly conditional on the chosen source model and common spectral assumptions. The observed domain-size sensitivity and unresolved discrepancy between the 3D response and the ideal planar benchmark prevent the interpretation of these coefficients as validated absolute heating rates. The next steps are to improve the convergence of the RF Hessian and weak mode, complete a systematic 3D domain and mesh response check, and connect the localized surface-potential model to physically justified or measured surface-noise statistics.

## AUTHOR DECLARATIONS

### Conflict of Interest

The authors have no conflicts to disclose.

## DATA AVAILABILITY

The data that support the findings of this study are available from the corresponding author upon reasonable request.